\documentclass[aps,prb,twocolumn,showpacs,showkeys,a4paper,floatfix]{revtex4-1}
\usepackage{subfig}
\usepackage{epsfig}
\usepackage{varioref}
\usepackage{rotating}
\usepackage{mathptmx}
\usepackage{fancyhdr}
\usepackage{amsmath}
\usepackage{amsfonts}
\usepackage{amssymb}
\usepackage{amsbsy}
\usepackage{textcomp}
\usepackage{array}
\usepackage{fixmath}
\usepackage{bm}
\usepackage[colorlinks]{hyperref}
\usepackage{eso-pic}

\begin{document}

\title{Electronic and optical properties of Cadmium fluoride: the role of many-body effects}
\author{Giancarlo Cappellini,$^{1,2}$ J\"urgen Furthm\"uller,$^3$ Emiliano Cadelano,$^{1,2}$ Friedhelm Bechstedt$^3$}
\email[Email:]{giancarlo.cappellini@dsf.unica.it}
\affiliation{$^1$Department of Physics, University of Cagliari, Cittadella Universitaria, I-09042 Monserrato (Cagliari), Italy \\
$^2$ Istituto Officina dei Materiali (IOM) del Consiglio Nazionale delle
Ricerche (CNR), Unita' Operativa SLACS, Cittadella Universitaria, I-09042 Monserrato (Cagliari), Italy\\
$^3$ETSF and IFTO, FSU-Jena, Max Wien Platz 1, D-07743 Jena, Germany}

\date{\today}

\begin{abstract}
Electronic excitations and optical spectra of $CdF_{2}$  are calculated
up to ultraviolet employing state-of-the-art techniques based on density functional theory and
many-body perturbation theory.
The GW scheme proposed by Hedin has been used for the electronic self-energy
to calculate single-particle excitation properties as energy bands and densities
of states.
For optical properties many-body effects, treated within the Bethe-Salpeter
equation framework, turn out to be crucial.
A bound exciton located about 1 eV below the quasiparticle gap is
predicted. Within the present scheme the optical absorption
spectra and other optical functions show an excellent agreement with
experimental data.
Moreover, we tested different schemes to obtain the best agreement with
experimental data.
Among the several schemes, we suggest a self-consistent quasiparticle energy
scheme. 
\end{abstract}
\maketitle

\section{Introduction}
\label{sec:introdution}
Fluorides and fluorite-type crystals have recently attracted much interest for
their intrinsic optical properties and their potential applications in
optoelectronic devices.\cite{rubloff} Because of their peculiar optical
properties, possible applications in ultraviolet (UV) laser optics could
be considered. For example, excimer ArF lasers with
an emission wavelength of 193 nm and KrF lasers with an emission wavelength
of 248 nm offer many applications in medicine: e.g., photoablation of the
substantia propria (or stroma of cornea), and high precision tissue
ablation.\cite{medicine application} This comes in addition to the common
application in photolithography of the semiconductor industry.

Here we will consider cadmium fluoride ($CdF_{2}$)  which is characterized
by a large gap energy (of the order of 9 eV)\cite{kalugin} and, hence, a high
transparency in a wide energy range, while the more famous calcium fluoride
($CaF_{2}$) has a direct band gap  at $\Gamma$ of $12.1$ eV and an
indirect band gap estimated around $11.8$ eV.\cite{rubloff} 
$CdF_{2}$ has been chosen as a
representative candidate for a material class which seems to
  exhibit similar behavior, as well as numerical problems  (e.g., $BaF_{2}$\cite{unpub}). 
  In particular, its overall very
  small values of the dielectric function, the extended high-energy tail, and the full ionicity
  (i.e., the weak band dispersions) may
  pose critical theoretical and numerical problems.

By doping, usually with trivalent metal impurities and a certain thermochemical
treatment, $CdF_{2}$ can be also produced in the semiconducting
state.\cite{hayes} Many dopants, as Sc and Y, generate only a shallow donor
state, but In or Ga related defects exhibit a bistable behavior. In addition to
the shallow state these impurity centers also possess a deep
state.\cite{piekara, dmochowski1,dmochowski2} 
Application of UV and visible light at low temperatures results in
disappearance of the absorption peak corresponding to the deep electronic state
and an infrared absorption band associated with the occupation of the
metastable shallow state. In other words, the electrons occupying the
deep state are lifted due to photoexcitation to the shallow state. 
At low temperatures the electrons cannot be trapped back to the deep states
due to a barrier induced by different atomic configurations for shallow and
deep states. The change in the electronic occupation causes a large difference
in the local refractive index, which can be used as means for holographic
writing at nanoscale spatial resolution.\cite{ryskin}
A general observation is that the formation energies of most defects are found
to be very low.\cite{mattila} This suggests that the fluorite structure in
$CdF_{2}$ often contains high defect concentrations in agreement to what is
found experimentally.\cite{hayes}
It is also essential that $CdF_{2}$ crystals of quite large size can be
obtained at a relatively low cost.\cite{kalugin}

Optical properties of $CdF_{2}$ have been experimentally determined by
different spectroscopic techniques since the seventies of the previous
century\cite{berger, bourdillon} in the fundamental absorption region and
in the core-level excitation range. In the same decade reflection spectra
of $CdF_{2}$ have been determined in comparison with those of
$SrF_{2}$,\cite{bourdillon} while UPS and XPS spectra of $CdF_{2}$ and
$SrF_{2}$  have appeared in the literature in 1980.\cite{raisin} Also
$\beta$-$PbF_{2}$ and $CdF_{2}$ mixed crystals absorption coefficients have
been reported by spectrophotometry measurements.\cite{kosacki}

Various theoretical methods have been applied to study either the ground state
or the excited states of the fluorite compounds. The energy bands and
reflectance spectra of $CaF_{2}$ and $CdF_{2}$ have been determined within a
combined tight-binding and pseudopotential method.\cite{albert}
Mixed crystals of $CaF_{2}$, $SrF_{2}$, $CdF_{2}$, $\beta $-$PbF_{2}$ have been
studied with respect to their electronic energy bands and density of states
(DOS) within the linear muffin-in orbital (LMTO) method.\cite{kudrnovsky}
Linear and non-linear optical properties of the cubic insulators $CaF_{2}$,
$SrF_{2}$, $CdF_{2}$, $BaF_{2}$ and other compounds have been determined by
first-principles orthogonalized linear combination of atomic orbitals (OLCAO).\cite{ching}
Point defect studies in $CdF_{2}$ have been performed within the plane wave
pseudopotential (PW-PP) method.\cite{mattila}
With respect to one-particle and two-particle electronic properties and energy
band gaps, state-of-the-art techniques have been applied until now only to
$CaF_{2}$. In fact, electronic band structures of $CaF_{2}$ have been
determined within a GW approximation, using a PW-PP scheme.\cite{shirley} 
On the other hand the imaginary part $Im ~\varepsilon (\omega)$ of the
dielectric function has been calculated for $CaF_{2}$ after an iterative
procedure using an effective Hamiltonian,\cite{benedict} within a PW-PP scheme
considering a screened interaction for electron-hole (\textit{e-h}) coupling,
and using localized orbitals.\cite{rohlfing} 

In this paper the electronic excitations and optical spectra of $CdF_2$ are
calculated up to 40 eV employing state-of-the-art techniques based on density functional
theory (DFT) and many-body perturbation theory (MBPT). We use the GW scheme
proposed by Hedin\cite{hedin} for the exchange-correlation (XC) self-energy,
employing an  Heyd-Scuseria-Ernzerhof (HSE)\cite{hse03} hybrid functional as starting point for the
electronic quasiparticle structure to calculate single-particle excitation
properties as the energy bands and the DOS. The DOS near the
gap region and the energy-band structure are compared with existing data from
literature; we show and discuss the comparison with them. The role of 
many-body effects turns out to be fundamental for these single-particle
properties. Moreover, we demonstrate that a simple one-shot GW ($G_{0}W_{0}$)
is not sufficient to obtain the correct band gaps.
For optical properties excitonic effects, treated within the Bethe-Salpeter
equation (BSE)\cite{onidarubioreining} framework, are crucial to allow a
reasonable comparison with existing experimental spectra as well. We discuss
the electronic excitation structure and the optical spectra including the
existence of bound excitons, and compare with available
experiments.\cite{kalugin}  
One of our goals is to provide a wide scenario of the one- and two-particle
properties of this material treated within first-principles techniques. 
However, as considered later on, a very extended high-energy tail and overall
rather small values of the dielectric function require inclusion of an
unusually large number of bands or band pairs in the GW and BSE schemes.
In regard to this last point, we suggest a theoretical framework and
 a simulative protocol able to give solutions with appreciable agreements with 
the available experimental measures, saving at the same time the computational demands. 

\section{Ground state properties and computational details}
\label{sec:Ground state properties}
All the calculations have been performed using density
functional theory (DFT)\cite{kohn} as implemented in the plane-wave basis code
VASP.\cite{kresse,kresse2} 
The projector augmented wave (PAW)\cite{paw,paw2} method is applied to generate 
pseudopotentials and wave functions in the spheres around the cores.
In standard use, VASP performs a fully relativistic calculation
for the core-electrons and treats valence electrons in a scalar
relativistic approximation.\cite{kresse,kresse2,bachelet,kaupp}
The effect of spin-orbit coupling on the energy bands has been considered
extensively in Ref. \onlinecite{cadelano}, where the spin-orbit effect, in
the case of $CdF_{2}$, slightly decreases the gaps by about 0.05 eV only.
Thus spin-orbit effects will be neglected in the present paper.

We performed the calculations using different \textit{XC} functionals. The local density
approximation (LDA) has been used for the \textit{XC} energy, as given by Ceperley and
Alder, and parametrized by Perdew and Zunger.\cite{ceperley,perdew} In
addition, the generalized-gradient approximation (GGA) in the parametrization
of Perdew, Burke, and Ernzerhof (PBE) is used.\cite{perdewburkeernzerhof}
Moreover, we considered also a revised version of the PBE functional which
gives  better results in solids which we shall refer to as
PBEsol\cite{perdew2008} from now on. 

The lattice structure of $CdF_{2}$, as well as of all the other fluorides, is
a cubic one with the space group \textit{$Fm\overline{3}m$}, with three ions
per unit cell, i.e., one cation $Cd$ placed in the origin and two anions $F$
situated at $\pm(\frac{1}{4}a, \frac{1}{4}a,\frac{1}{4}a)$.\cite{wyckoff}
In the crystal the  $F^{(-)}$ ions form a simple cubic sublattice surrounded
by a face-centered cubic lattice of $Cd^{(++)}$ cations.
All fluorides with cations belonging to the II and IIB groups are stable in
this crystallographic structure.\cite{cadelano}

Besides the $Cd$(5\textit{s}) and $F$(2\textit{s},2\textit{p}) valence states also
the shallow $Cd$(4\textit{s}, 4\textit{p}, 4\textit{d}) core states are treated as valence electrons.
Inclusion of the $Cd$(4\textit{d}) states is important because they appear in between the
$F$(2\textit{s}) and $F$(2\textit{p}) bands. Inclusion of the strongly bound $Cd$(4\textit{s},4\textit{p}) states (about
100 or 60 eV below the valence band maximum) is not an absolute must but
helps a lot to construct \textit{s}- and \textit{p}-pseudopotentials with excellent scattering
properties up to very high energies which are needed for a proper treatment of a
large number of unoccupied bands in the calculation of the dielectric function.
The wave functions are expanded in a plane-wave basis set with a cutoff energy
of 950 eV. The DFT/HSE calculations for this cutoff value are fully converged. For GW/BSE calculations 
the error bar has been estimated at about 0.1 eV.
The face-centered cubic (fcc) Brillouin zone (BZ) is sampled by $\Gamma$-point
centered meshes with 16\texttimes 16\texttimes 16 Monkhorst-Pack
\textit{k}-points\cite{monkhorst} (converging within a few meV).
The minimum of the total energy with respect to the volume is
obtained by fitting to the Vinet equation of state.\cite{vinet}
For the calculation of the cohesive energies $E_{coh}$, we have subtracted
the spin-polarized ground-state energies of the free atoms.
\begin{table}[bp]
\caption{Structural data for $CdF_{2}$ are given: lattice constant $a_{\circ}$, bulk modulus $B_{\circ}$ and its pressure derivative $B'_{\circ}=dB_{\circ}/dP$, and cohesive energy $E_{coh}$. Three local and semilocal \textit{XC} functionals are used, as discussed in the text. Other theoretical results  and experimental references are reported in column ''OTheory'' and in column ''Expt.'', respectively.}

\label{table: lattice constant}
\centering
 \begin{tabular}{l|ccccc}
\hline
\hline
	      	  &LDA		& PBE 		& PBEsol	& OTheory 			&  Expt.\\
\hline
$a_{\circ}$ (\AA)  &5.30		& 5.49		& 5.39	 	& 5.39\cite{kudrnovsky}	        & 5.36\textminus5.39\cite{deligoz}\\
$B_{\circ}$ (GPa)  &126.70 	&93.80		& 108.50 	& 123.00\cite{kudrnovsky}	& 114.60\cite{deligoz}\\
$B'_{\circ}$      &   4.76  	& 4.88  	&  4.85 	& 4.85\cite{kudrnovsky} 	& ~ \\
$E_{coh}$ (eV)    & -12.07 	& -9.55		&  -10.39 	& ~	& ~ \\
\hline
\hline
\end{tabular}
\end{table}
Comparing experimental measurements\cite{kudrnovsky} with results calculated
with different \textit{XC} functionals, see Table \ref{table: lattice constant},
the best results are obtained within the PBEsol scheme. Therefore we have
decided to use from now on the PBEsol lattice constant of $5.39$ \AA~
consistently for all calculations (ground-state, one- and two-particles
excitations).
This is an important point since band structure test calculations on DFT
level for different lattice constants show that the gap value changes by about
+0.1 eV (-0.1 eV) when going from the PBEsol to the LDA (PBE) lattice constant.

\section{Quasi-particle excitations}
\label{sec:computational}
The Kohn-Sham (KS) eigenvalues of the DFT scheme cannot be interpreted as
energies of single-particle/quasiparticle (QP) electronic excitations and,
therefore, cannot be compared with the band structure and the DOS from
experiments.
We apply the many-body perturbation theory within Hedin's GW approximation
for the electronic self-energy operator.\cite{hedin}
It can be demonstrated that one has to solve the QP
equation:\cite{hl,onidarubioreining,cappellinissc92,Stankovski}
\begin{eqnarray}
 \label{eq:1}
\nonumber
 \left\{ -\frac{\hslash^{2}}{2m_{o}}\nabla_{\mathbf{r}}^{2}+\nu_{H}(\mathbf{r})+\nu_{ext}(\mathbf{r}) \right\}\Phi_k^{QP}(\mathbf{r})\\
+\int d^{3}\mathbf{r'}\varSigma(\mathbf{r},\mathbf{r'};\varepsilon_k^{QP})\Phi_k^{QP}(\mathbf{r'})=\varepsilon_k^{QP}\Phi_k^{QP}(\mathbf{r}).
\end{eqnarray}
Equation (\ref{eq:1}) has a similar structure as the KS equation of DFT. The
fundamental difference is that the \textit{XC} potential of DFT is replaced by the
non-Hermitian, nonlocal energy dependent self-energy operator $\varSigma$
resulting in QP eigenvalues $\epsilon_k^{QP}$ and QP wave functions
$\Phi_k^{QP}$. 
Due to the fact that the Hamiltonian is non-Hermitian, QP eigenvalues are
usually complex numbers. The real part describes resonance energies
(excitation energies) and the imaginary part defines the lifetime of
excited states. 
Equation (\ref{eq:1}) should be solved self-consistently;
however, the most common GW approach avoids this procedure.
Since empirically one often finds that QP wave functions are similar to the
KS DFT-LDA ones,\cite{hl} it is natural to take them as a starting point for a
GW calculation and, hence, to calculate QP corrections in the sense of a
perturbation treated within first-order perturbation
theory.\cite{hl,onidarubioreining,cappellinissc92}
The following equation,
\begin{equation}
 \label{eq:2}
\varepsilon_k^{QP}=\varepsilon_k+\langle  \varphi_{k} \lvert \varSigma(\epsilon_k^{QP})-\nu_{XC} \rvert \varphi_{k} \rangle,
\end{equation}
is used in conjunction with the random phase approximation (RPA) dielectric
function to calculate the screened Coulomb interaction $W_0$ and the DFT
Green's function $G_0$. This approach called $G_{0}W_{0}$ usually works very
well and gives very good results for the GW eigenvalues (band gaps, bandwidths
and band dispersions).\cite{hl, cappellinissc92} 
Another GW treatment of the QP band structure is based on the so-called
generalized KS schemes (gKS).\cite{gKS1,gKS2}
The gKS methods have in common that nonlocal exchange-correlation potentials
involving partial or screened exact exchange are inserted into the KS equations
combined with (semi-)local DFT potentials. From the viewpoint of band structure
calculations these gKS schemes can be also considered as oversimplified
GW schemes. Hence, they can be considered as an improved starting point
(eigenvalues, wave functions) for $G_{0}W_{0}$ calculations. While results are
similar for many systems, gKS schemes as a $G_{0}W_{0}$ starting point can
provide much improved results for those cases where DFT-LDA or -GGA fails
drastically (wrong band ordering, extreme gap underestimates). Since the
experimental gap is almost three times as large as the DFT-PBEsol one (a gap
correction of about 5 eV is necessary to arrive at the experimental values) a
gKS starting point (still leaving a 3 eV gap error) can be considered as an
improvement also for fluorides.
\begin{figure}[tp]
\includegraphics[width= 0.50\textwidth, angle=0]{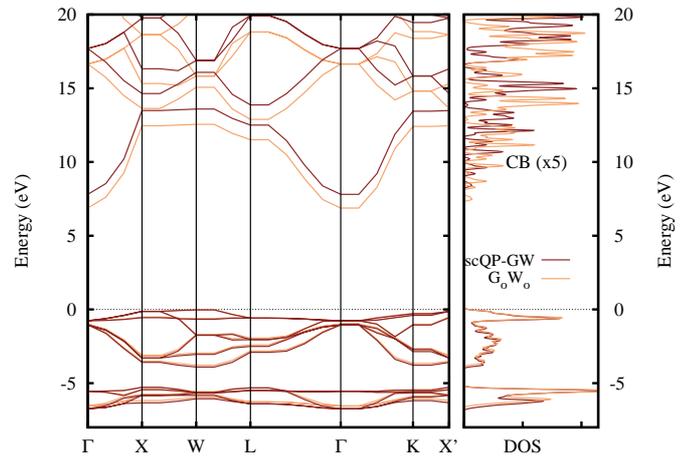}
\caption{\label{fig:1}(Color online) Energy bands of $CdF_{2}$ calculated within the $G_{0}W_{0}$ (light lines) and the scQP-GW (dark lines) schemes. On the right panel of the figure, the DOS are reported for valence bands (VB) and conduction bands (CB). Notice the magnification factor used for the conduction bands.}
\end{figure}
In the present work QP bands of $CdF_{2}$ are determined by an iterative
solution of the QP equations given in Eq.(\ref{eq:2}) with the \textit{XC} self-energy
in the GW approximation. The iteration starts with the gKS equations, with
a self-energy derived from the nonlocal HSE03 hybrid functional
(zeroth order).\cite{hse03}
Subsequently, the GW corrections are calculated in first-order
perturbation theory, i.e., within the one-shot $G_{0}W_{0}$.
Thereby, the full frequency dependence of the RPA dielectric function entering
the screened Coulomb interaction ($W_0$) is taken into account sampled on a
frequency grid with 128 points up to $\hbar\omega\approx$350 eV, i.e., no
plasmon pole models\cite{Stankovski} or model dielectric functions\cite{cappellinissc92} have been involved.
We present here QP eigenvalues calculated in the $G_{0}W_{0}$ 
approach on top of the HSE03 ground-state electronic structures.

The strategy to calculate QP corrections then follows: The band structure and
DOS have been calculated first with the PBEsol GGA, then a HSE03 calculation is
performed, and finally we apply the $G_{0}W_{0}$ scheme on top of HSE03.
Even after this step, the QP energies are of the order of $1$ eV smaller than
the experimental ones. For comparison we made also Hartree-Fock (HF)
calculations, which show as expected deviations in the opposite direction
(i.e., an unrealistic overestimate of the gaps).\cite{hl, strinati}
As the last step, in order to open that gap further, we performed a
''self-consistent quasiparticle energy'' calculation within the GW
approximation (namely, scQP-GW) which is a kind of ''iterated $G_{0}W_{0}$'' 
just updating the eigenvalues only but keeping the HSE03 wave functions fixed.
This procedure gives almost perfect gaps with respect to experiment.\cite{smith, caruso}

\begin{figure}[tp]
\includegraphics[width= 0.50\textwidth, angle=0]{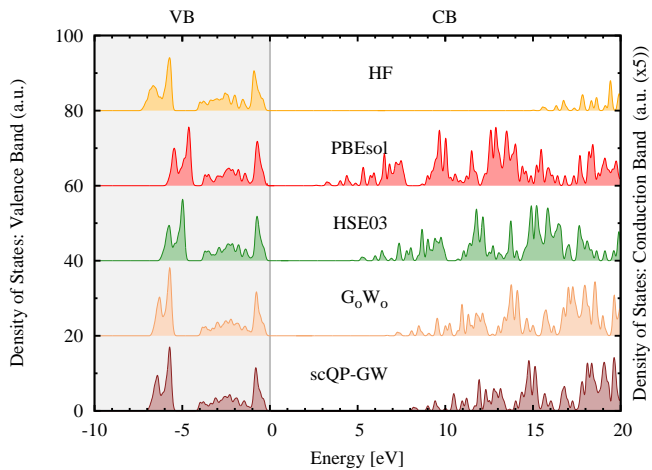}
\caption{\label{fig:2}(Color online) Density of states around the gap region
calculated within different approximations: HF, PBEsol, HSE03, $G_{0}W_{0}$, and
scQP-GW. Notice the magnification factor (five) used for the conduction bands.}
\end{figure}

The main outcomes of the present paragraph are reported in Table \ref{table:2} and
in Figs. \ref{fig:1} and \ref{fig:2}. More details are also discussed in Ref. \onlinecite{note2}.
The QP values for the fundamental gaps $\Gamma-\Gamma$ (direct) and $W-\Gamma$
(indirect) obtained within different approximations are also reported in
Table \ref{table:2}.
\begin{table}[bp]
\caption{QP energies for the fundamental gaps of $CdF_{2}$, between lowest CB and highest VB at high symmetry points, are computed with
several approximations and compared with available experiments.}
\label{table:2}
\centering
\begin{tabular}{c|cccccc}
\hline
\hline
           data in eV         &HF   &PBEsol & HSE03 &  $G_{0}W_{0}$ & scQP-GW &  Exp.\\
\hline
$\Gamma-\Gamma$ & 16.02  & 3.41 &5.50 & 7.64  &  8.58 & 8.4\cite{orlowski},8.7\cite{raisin} \\
$W-\Gamma$     & 15.30  & 2.91 & 4.91 & 6.90 & 7.83 & 7.8\cite{orlowski} \\
\hline
\hline
\end{tabular}
\end{table}
\begin{table}[bp]
\caption{\label{tab:mygap}Transition energies between lowest CB and highest VB
at high symmetric points, and valence bandwidth within the two GW schemes. All
data are expressed in eV.}
\centering
\begin{tabular}{c|cc}
\hline
\hline
Direct bandgaps     	 & $G_{0}W_{0}$   &  scQP-GW  \\
\hline 
$L\rightarrow L$ 	 &  12.11      &  13.08     \\
$\Gamma\rightarrow\Gamma$&   7.64      &  8.58               \\
$ X\rightarrow X $       &  12.57      & 13.62                \\
$ W\rightarrow W $       &  12.28      & 13.66                \\
$ K \rightarrow K $      &  12.65      & 13.44                \\
\hline
Indirect bandgaps 	 &       &                          \\
\hline
$W\rightarrow\Gamma $	 & 6.90  &   7.83                       \\
$W\rightarrow X $  	 & 12.50 &   13.55               \\
$W\rightarrow L  $  	 & 11.56 &   12.54                       \\
$W\rightarrow K $  	 & 12.47 &   13.44                       \\
\hline
VB width (F 2\textit{s} - F 2\textit{p}) &  22.47    &  22.99                       \\
\hline
\hline
\end{tabular}
\end{table}
It is clear from Table \ref{table:2} and from Figs. \ref{fig:1} and \ref{fig:2}
that after the computational procedure adopted here an excellent agreement
has been obtained with available experiments.\cite{orlowski,raisin} The other
point is that for $CdF_{2}$, $G_{0}W_{0}$ is not sufficient and a
self-consistent procedure, GW based, has to be considered in addition.
Referring to Fig. \ref{fig:1} the main discrepancies between the different GW
methods appear in the conduction bands. This point is also clear from the DOSs
which are nearly overlapping for the VB energy region, as shown in
Figs. \ref{fig:1} and \ref{fig:2}. 

Therefore, the electronic structure of $CdF_{2}$, within the PBEsol,
$G_{0}W_{0}$, and scQP-GW schemes, shows an indirect gap at $W-\Gamma$ of
7.83 eV (expt. 7.8 eV\cite{orlowski}) and a larger direct gap at $\Gamma$ of
8.58 eV (expt. $8.4-8.7$ eV).\cite{orlowski,raisin} The best accordance with
experiments is found within the scQP-GW scheme with a high level of accuracy.
In Fig. \ref{fig:2} we present the calculated DOS within different schemes.
As is clear from this figure the VB regions after the HSE03 and PBEsol methods
seem very similar while discrepancies arise for the CB region between the
results after the two methods (higher value of the gap after the HSE03 scheme).
 
Similarities for the VB regions  appear also between the result after the
$G_{0}W_{0}$ and scQP-GW schemes. The latter method does operate a further
opening of the gap. In the upper part of Fig. \ref{fig:2}, the HF DOS is
reported. It determines a (too) large overestimate of the band gaps (about
16 eV at $\Gamma$) and valence bands are only partially (the shallow ones)
similar to those calculated after the other methods.
This HF band gap overestimate resembles a similar behavior taking place in
other systems.\cite{hl, strinati}
Notice that the \textit{d}-electrons of $Cd$ play an important role. Indeed, they are
present inside the valence bands in the energy range displayed in
Fig. \ref{fig:1}, and give rise to flat bands/sharp peaks near -5 eV, due to the
localized character of these states. Localization of CB states is much weaker,
and hence their dispersion is much stronger. Therefore, the average height of
the DOS is smaller in comparison with the VB states.
It is interesting to report in detail the energy band gaps
calculated within the two different GW schemes proposed in the
present publication, namely $G_{0}W_{0}$ and scQP-GW.
As it is clear from Table \ref{tab:mygap} a nearly rigid shift of about 1 eV
appears going from $G_{0}W_{0}$ to scQP-GW transition energies.
Therefore, with respect to $CdF_{2}$, considering Fig. \ref{fig:1} and
Table \ref{tab:mygap}, the action of the self-consistency upon the eigenvalues
operates a kind of ''second order scissor operator rule'' on the conduction
bands to obtain good accordance with experiments.
The valence bandwidths after the two approximations of \textit{XC}, respectively
$G_{0}W_{0}$ and scQP-GW, agree within 0.5 eV. Thereby, the scQP-GW value of
22.99 eV corresponds excellently to an experimental estimate of 23 eV reported
by Raisin.\cite{raisin}

\section{Two-particles properties and dielectric function}
\label{sec:2particle}
The treatment of excitons and resulting optical spectra, i.e., the full
excitonic problem, requires to set up and diagonalize an electron-hole
Hamiltonian \^{H}. Within Hedin's GW scheme and a restriction to static
screening this two-particle Hamiltonian reads for singlet excitations in
matrix form as:\cite{onidarubioreining,cappellini2}
\begin{eqnarray}
 \label{eq:3}
\nonumber
{\hat H}(v c\mathbf{k},v' c'\mathbf{k'})=[\varepsilon_c^{QP}(\mathbf{k})-\varepsilon_v^{QP}(\mathbf{k})]\delta_{v v'}\delta_{c c'}\delta_{\mathbf{k} \mathbf{k'}} \\
\nonumber
-\int d^{3}\mathbf{r} \int d^{3}\mathbf{r'} \varphi_{c\mathbf{k}}^{\ast}(\mathbf{r})\varphi_{c'\mathbf{k'}}(\mathbf{r})W(\mathbf{r},\mathbf{r'})\varphi_{v\mathbf{k}}(\mathbf{r'})\varphi_{v'\mathbf{k'}}^{\ast}(\mathbf{r'})\\
+2\int  d^{3}\mathbf{r}\int d^{3}\mathbf{r'} \varphi_{c\mathbf{k}}^{\ast}(\mathbf{r})\varphi_{v\mathbf{k}}(\mathbf{r}) \bar{v}(\mathbf{r},\mathbf{r'})\varphi_{c'\mathbf{k'}}(\mathbf{r'})\varphi_{v'\mathbf{k'}}^{\ast}(\mathbf{r'}),
\end{eqnarray}
where matrix elements between KS or gKS wave functions of CB states and VB
states occur. Contributions to Eq.(\ref{eq:3}) which destroy particle number
conservation have been omitted.
The first term describes the noninteracting quasi-electron quasi-hole pairs. 
The second term accounts for the screened electron-hole Coulomb attraction with
the statically screened Coulomb potential $W(\mathbf{r},\mathbf{r'})$.
The third contribution, governed by the nonsingular part of the bare Coulomb
interaction $\bar{v}(\mathbf{r},\mathbf{r'})$, represents the electron-hole
exchange or crystal local-field effects.\cite{onidarubioreining,roedl}
The QP eigenvalues entering the first term and $W(\mathbf{r},\mathbf{r'})$
entering the second term have been directly extracted from the scQP-GW results.
After diagonalization of the exciton matrix, or, more precisely, solving the
homogeneous BSE or stationary two-particle Schr\"odinger equation with the
eigenvalues $E_{\varLambda}$ and the eigenfunctions
$A_{\varLambda}(v c\mathbf{k})$ of the pair states $\Lambda$,
frequency dependent macroscopic dielectric function including excitonic effects
can be written as:
\begin{eqnarray}
 \label{eq:4}
\nonumber
\varepsilon_{\alpha\alpha}(\omega)&=&\delta_{\alpha\alpha}+\frac{16\pi e^{2}\hslash^{2}}{V}\sum_{\varLambda}\left\lvert \sum_{cv\mathbf{k}}\frac{\langle c\mathbf{k} \lvert v_{\alpha} \rvert v\mathbf{k}\rangle}{\varepsilon_c(\mathbf{k})-\varepsilon_v(\mathbf{k})}A_{\varLambda}(v c\mathbf{k}) \right\rvert^{2}\\
&\times& \left[ \frac{1}{E_{\varLambda}-\hslash(\omega+i\gamma)}+\frac{1}{E_{\varLambda}+\hslash(\omega+i\gamma)}  \right] 
\end{eqnarray}
where $v_{\alpha}$ is the corresponding Cartesian component of the
single-particle velocity operator and  $\gamma$ is the pair damping constant.
The crystal volume is given by V. The details of the standard scheme using the
direct diagonalization of Eq.(\ref{eq:4}) have been discussed
elsewhere.\cite{onidarubioreining, shirley,cappellini2,roedl}

Since the rank of the Hamilton matrix in Eq.(\ref{eq:4}) is extremely large (it is given by the number of
valence bands times the number of conduction bands times the number of
$\bf k$ points) a straightforward diagonalization of this
matrix is often not possible due to high CPU but also memory requirements.

Therefore, we mainly refer to a numerically efficient
scheme\cite{smith} to solve the Bethe-Salpeter equation, which has been
successfully applied by Bechstedt and coworkers on several systems.
This method is based on the calculation of the time evolution of the
exciton-state by Fourier transform on the time domain. This scheme delivers
optical spectra directly but no exciton eigenvalues and
eigenfunctions.\cite{smith} 
The calculation of the $\omega$-dependent polarizability can be considered
as an initial-value problem.\cite{smith} This means that it can be written
as a Fourier representation within an integral of time-dependent elements.
The time evolution of these elements is driven by the pair Hamiltonian [see
Eq.(\ref{eq:3})] and it is operatively performed  by using a central-difference
method, which requires one matrix-vector multiplication per time step. The time
integration can be truncated due to exponential decay factors and it determines
a scheme nearly independent from the dimension \textit{N} (number of pair states) of the
system. Two further advantages come from the matrix-vector multiplication
scheme, i.e., an $O(N^2)$ dependence on operations count and the
possibility to effectively distribute the multiplications on several processors
of a parallel computer.\cite{smith}
Applications of this method to the calculation of the $\omega$-dependent
dielectric function of bulk and surface systems have been successfully
performed with good comparison either with experimental results and with
outcomes of the matrix diagonalization scheme.
Moreover, a further speed-up of the calculations could be obtained combining
this method with the use of a model dielectric
function,\cite{smith, cappellinissc92, cappelliniPRB93} which, however, has
not been employed in the current calculations which involve no models at all.
This scheme has the big advantage that it never requires the explicit
diagonalization of the Hamilton matrix, because it needs only to evaluate the
Hamilton exciton matrix  times the current wave function for every time-step.
Massive CPU-time saving  and improved scaling of the computational work with
respect to system size result. On the other hand, we should admit that the lack of eigenvalues and eigenfunctions limits the detailed  knowledge of excitons, e.g., their real-space representation.\cite{note4}
\begin{figure}[tp]
\includegraphics[width= 0.50\textwidth, angle=0]{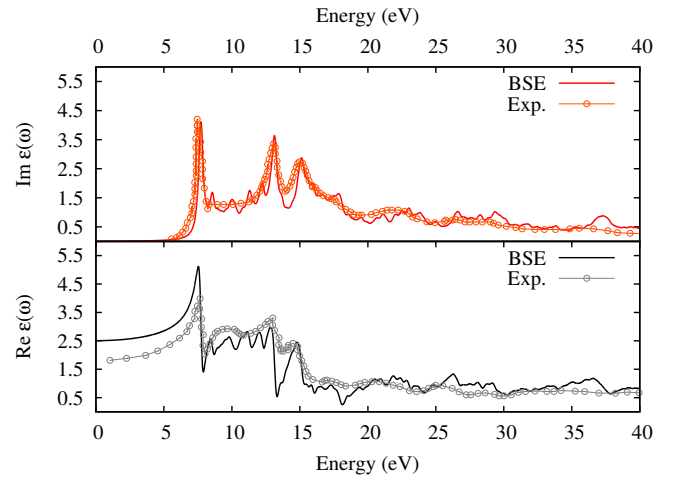}
\caption{\label{fig:3}(Color online) Imaginary and real part of the dielectric function calculated within the BSE scheme in comparison with experimental results after Ref. \onlinecite{bourdillon}.}
\end{figure}
The investigation of the fine structure of optical absorption
spectra, especially near the absorption edge and for low energy
optical transitions, requires finer \textit{k}-point samplings
than those necessary for the ground-state calculations.
Due to the large number of bands and high cutoffs involved we have, however,
to restrict ourselves to a rather limited number of \textit{k}-points.
For the GW and BSE calculations, we apply a
$\Gamma$-centered 8\texttimes8\texttimes8 mesh.
Furthermore, a sufficiently large number of conduction
bands has to be included in the calculations to describe the
electron-hole pair interaction properly. The number of empty
states is limited by introducing a cutoff energy for the electron-hole
single-particle transitions (without QP corrections) that contribute to the
excitonic Hamiltonian of Eq.(\ref{eq:3}). This transition energy cutoff was
chosen to be 60 eV, corresponding to up to 50 unoccupied bands. A rank of the
excitonic Hamiltonian matrix of the order of 210.000 results from this setup.
In addition, for the calculation of the full frequency dependent dielectric
function, screened exchange integrals, and four-orbit integrals in the GW and
BSE schemes, one may introduce a lower plane-wave cutoff than for ground-state
calculations without spoiling accuracy too much. In our case we use a cutoff
of 475 eV.

The matrix elements of the excitonic Hamiltonian have
been evaluated by means of HSE03 wave functions. Since
the  HSE03+$G_{0}W_{0}$  QP eigenvalues still leads to a gap
underestimate (see Table \ref{table:2}), HSE03+$scQPGW$  eigenvalues
are employed instead of HSE03+$G_{0}W_{0}$ eigenvalues on the main
diagonal of the Hamiltonian. It has been carefully checked that
no change of band ordering occurs when going from PBEsol to
HSE03. 

\begin{figure}[tp]
\includegraphics[width= 0.50\textwidth, angle=0]{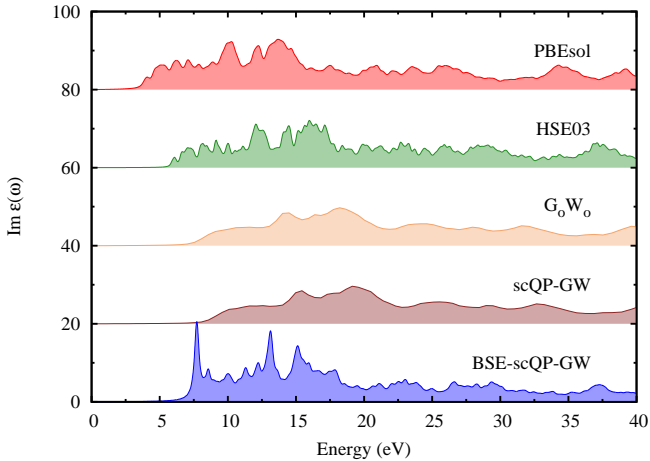}
\caption{\label{fig:4}(Color online) Imaginary part of the macroscopic dielectric function calculated within different approximations, namely PBEsol, HSE03, $G_{0}W_{0}$, scQP-GW, and BSE. Notice that each graph has been shifted by 20 units along the \textit{y} axis.}
\end{figure}
In Fig. \ref{fig:3}, we report the principal result of the present
calculations: the imaginary part of the dielectric function
 $Im ~\varepsilon (\omega)$ determined within the BSE approach by using the method
of Schmidt \textit{et al.}.\cite{smith} The real part has been obtained by a
Kramers-Kronig (KK) transform afterwards.
Also the experimental results after Bourdillon\cite{bourdillon} are reported,
which are obtained with synchrotron radiation technique. The agreement both in
the positions and intensities of the peaks between theory and experiment should
be stressed for the imaginary part of the dielectric function (e.g., the first main
three peaks are located at 7.7, 13.1, and 15.1 eV, respectively). As shown in
the top panel of Fig. \ref{fig:3}, the good agreement turns out for the full
energy spectrum from near UV up to 40 eV, either below and above the QP direct
gap.
The first structure is a $\Gamma$ exciton, while the second and the
third peaks are excitons which may originate from \textit{X}, \textit{L}, or \textit{W} transitions
between valence and conduction bands. Since we calculated the spectra directly
by the method of Schmidt, we have no access to excitonic eigenvalues
or eigenfunctions and can just speculate on the basis of single-QP eigenvalues.

However, on the basis of the experimental spectra of
Bourdillon\cite{bourdillon} as shown in Fig. \ref{fig:3}, the value
of $\varepsilon_{\infty}$ is seemingly about 1.8, which is in large
discrepancy to our value of 2.5. However, one should mention that the data of
Bourdillon have been extracted indirectly from reflection spectra and that
there exist other values for $\varepsilon_{\infty}$ from more direct methods.
Most probably, the best directly measured value is about 2.4,\cite{bosomworth}
and it is also indirectly confirmed by Krukowska\cite{krukowska} from
measurements of the refraction index at visible optical frequencies, obtaining
a value slightly larger than 2.4, i.e., about 2.46\textminus2.5. Furthermore, by
using the Lyddane-Sachs-Teller (LST) relation\cite{fox} the consistency of our
result could also be indirectly confirmed from measurements of $\epsilon_0$ 
(which yield values of 7.7, 7.8, and 8.3\textminus8.5  at 0 K,  80 K, and 300 K,
respectively)\cite{young} together with phonon data,\cite{deligoz, cribier}
which lead toward values in the range of 2.3\textminus2.5 for $\varepsilon_{\infty}$.
Hence all these values from direct measurements are in excellent agreement with
our computed value.
Furthermore the real part of the dielectric function $Re ~\varepsilon (\omega)$,
as shown in the second panel of Fig. \ref{fig:3}, is in close agreement with the
experimental data after Ref. \onlinecite{bourdillon}.

In Fig. \ref{fig:4}, we show the imaginary part of the dielectric function
$Im ~\varepsilon (\omega)$ within the different approximations  proposed here,
starting from the DFT reference which is the PBEsol one. Going from this
approximation to the gKS HSE03 one an opening of the gap results. Both GW
schemes further enlarge the gap energies. Finally the BSE results are disposed
which reproduce the experiment at best in terms of oscillator strengths,
peak positions, and onset behavior.

\section{Optical and loss functions}
\label{sec:optical}
\begin{figure}[tp]
\includegraphics[width= 0.50\textwidth, angle=0]{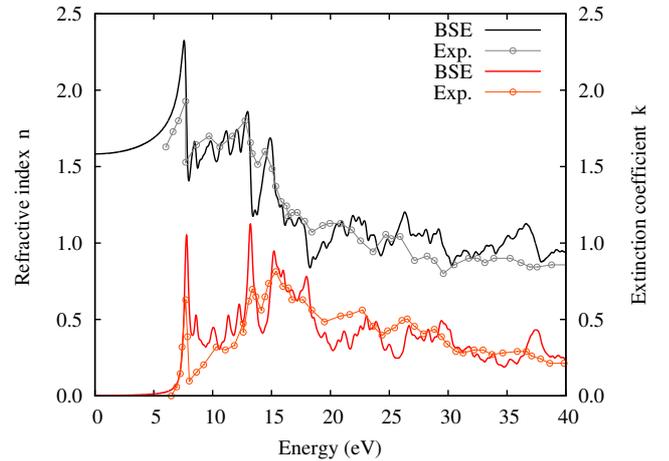}
\caption{\label{fig:5}(Color online) Real and imaginary part of the refractive index (\textit{n},\textit{k})
 as a function of energy in comparison with measured spectra (after  Ref. \onlinecite{bourdillon}).}
\end{figure}
In Fig. \ref{fig:5}, the imaginary and real part of the refractive index in
comparison with experiment after Ref. \onlinecite{kalugin} are shown.
A fair agreement results with respect to the three main
peaks at 7.8, 13.2, and 15.2 eV and to the high-energy region 20\textminus40 eV.

\begin{figure}[tp]
\includegraphics[width= 0.50\textwidth, angle=0]{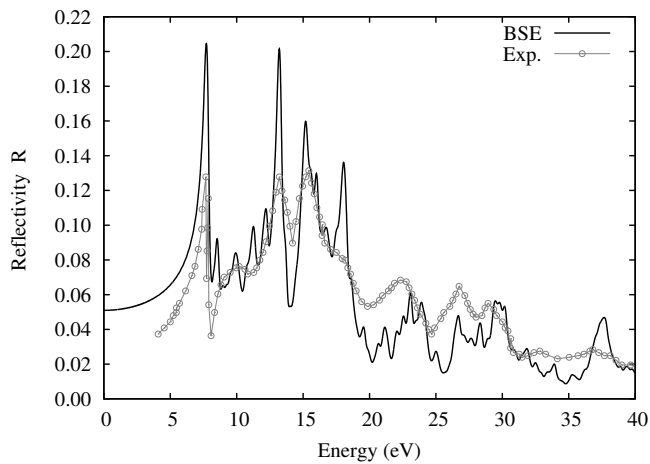}
\caption{\label{fig:6}Calculated reflectance spectrum within BSE and experiments after Ref. \onlinecite{bourdillon} as a function of energy.}
\end{figure}

The reflectivity function $R$ is show in Fig. \ref{fig:6} in comparison with an
experimental spectrum\cite{bourdillon}. The positions of the main peaks of the
theoretical curve (namely, occurring at 7.8 eV, 13.2 eV, and 15.2 eV) reproduce
the corresponding ones of the measurement, except for the fourth peak (at
18.0 eV) which is overestimated in intensity and it should be compared with the
shoulder around 17 eV as obtained in the experimental spectrum. Furthermore, the
energy region above 16 eV is only in qualitative agreement with experiment.
Regarding the first three peak positions of the reflectivity \textit{R}, our results compare well with the
experimental values after Berger and coworkers.\cite{berger}
Albert \textit{et al.}\cite{albert} assign these peaks to
$\Gamma_{15}^{V}\rightarrow \Gamma_{1}^{C}$, $X_{2'}^{V}\rightarrow X_{3}^{C}$
and $X_{5'}^{V}\rightarrow X_{3}^{C}$ excitons.
The first excitonic peak at 7.7 eV is consistent with the results of 7.6 eV given by
Orlowski\cite{orlowski}, Raisin\cite{raisin}, and by Forman\cite{forman}
and coworkers.

We focus on the fact that in our computations a natural/instrumental
broadening of 0.2 eV has been applied in all the curves.
Within our scheme, ''our experimental curve'' is the imaginary part of the
dielectric function, $Im ~\varepsilon (\omega)$ (the first outcome of the
computational code). All the other functions are derived from this calculation.
Thus, an important methodological point is the following: From each optical function
produced here one can return to the original dielectric function which was the
output of the code we used (''our reference result''). Moreover the comparison
with the experimental result for $Im ~\varepsilon (\omega)$ is anyway excellent
(see Fig. \ref{fig:3}). If one operates a broadening to fit at best an optical
function with respect to a particular experiment, the reverse operation (to
generate again the $Im ~\varepsilon (\omega)$ in agreement with experiment) is
not guaranteed.

\begin{figure}[tp]
\includegraphics[width= 0.50\textwidth, angle=0]{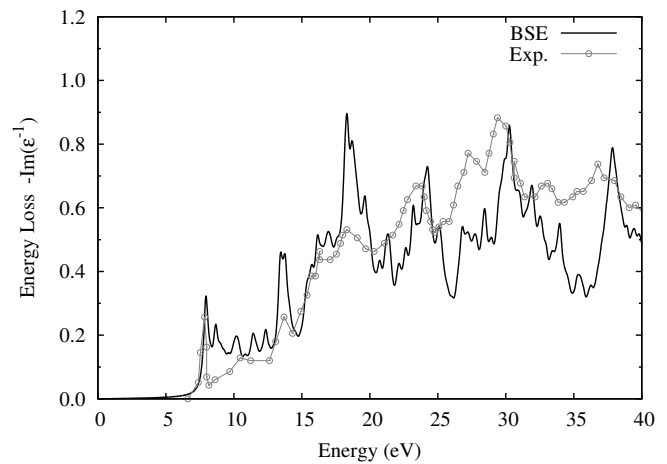}
\caption{\label{fig:7}Electronic energy loss function after present calculation vs experiments (Ref. \onlinecite{kalugin}).}
\end{figure}

In Fig. \ref{fig:7}, we present the energy loss function after the present BSE
calculation and experiment in a wide energy range.\cite{kalugin}
Also for the onset region and the main three structures below 20 eV the
agreement is very close in positions and, a little less, in the oscillator
strengths. For energies larger than 20 eV, a slight blue-shift of the peak
positions is obtained with respect to the experimental ones.
Furthermore, some differences with respect to the energy loss function
$-Im ~\varepsilon^{-1}$ of other semiconductor/insulator systems could be
considered.
Indeed, these systems generally show a broad single structure around the
plasmon frequency $\omega_p$ (e.g., as in the cases of cubic GaAs and Si),\cite{cardona} unlike our theoretical curve (as shown
in Fig. \ref{fig:7}), while a structure of well-separated peaks appears in the spectrum of $CdF_2$. 
A single broad structure would appear
around the energy of 34.6 eV, if all the valence electrons had contributed to a
single plasmonlike excitation. Instead of this hypothetic structure, only two sharp peaks (located at 7.9 and 13.4 eV, respectively) could be ascribed to excitonic effects.
Analyzing the positions of the peaks above 15 eV, it seems that they
occur around the energies for which the real part of the dielectric function (see Fig. \ref{fig:3}) shows its minima; i.e., around 19 eV, 25 eV, 30 eV, and 37 eV.
In particular, one could suggest that due to the distinct separation of the
valence states, in the case of $CdF_2$ the plasmon excitations take place
but divided in well-defined energy windows.
One could argue that the electrons remain at disposal for collective
plasmonlike excitations as far as the corresponding energies are larger than
the relative binding energies.
Nevertheless, this hypothesis is correct for the  $F$(2\textit{p}), $Cd$(4\textit{d}), and $F$(2\textit{s})
valence electrons but it is no more true for the $Cd$(4\textit{p}) and $Cd$(4\textit{s}) electrons.   
In fact, considering the contribution of the $F$(2\textit{p}) electrons one obtains a
plasmon frequency of 20.53 eV.  The contribution of $F$(2\textit{p}) plus $Cd$(4\textit{d}) electrons
gives a plasmon energy of 27.81 eV. Finally, a plasmon energy of 30.23 eV is
obtained including the $F$(2\textit{s}) electrons. These values are not so far away from
the energies of the first three above-mentioned peaks of loss function.
On the other hand, the $Cd$(4\textit{p}) and $Cd$(4\textit{s}) electrons lead to the following
values for collective plasmonlike excitations: 32.53 eV [Cd(4\textit{p} included]  and
34.57 eV [Cd(4\textit{p}) and $Cd$(4\textit{s}) included]. These energies are much smaller than the
corresponding binding energies of about -62 eV for $Cd$(4\textit{p}) and -102 eV for $Cd$(4\textit{s})
electrons. Therefore, a plasmonlike mechanism,
involving $Cd$(4\textit{p}) and $Cd$(4\textit{s}) electrons, could not explain the structures in the
high-energy spectrum of Fig. \ref{fig:7}. It is interesting to note that the
electron energy loss function of the rock salt rs-CdO, similarly to the present
case, shows no evidence of a single pronounced plasma resonance coming from
\textit{s}, \textit{p} or \textit{d}  electrons.\cite{schleife} 

\begin{figure}[tp]
\includegraphics[width= 0.50\textwidth, angle=0]{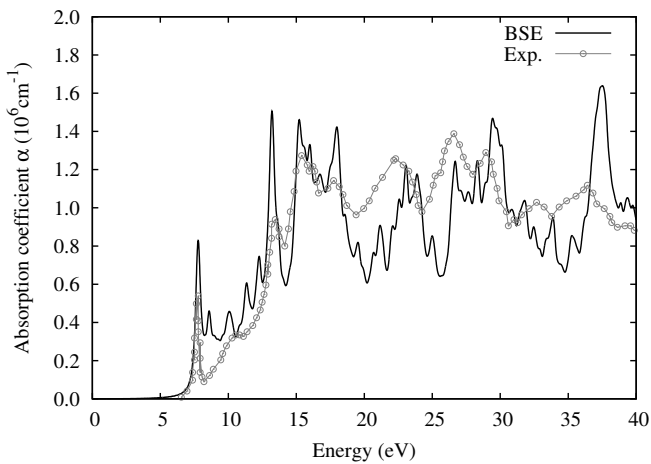}
\caption{\label{fig:8}Absorption coefficient after present calculation vs experiment (Ref. \onlinecite{kalugin}).}
\end{figure}

In Fig. \ref{fig:8}, the absorption coefficient $\alpha$ is shown in comparison
with available experimental data.\cite{kalugin} 
Considering the first part of the $\alpha$ function, the main four sharp peaks
(at 7.8 eV, 13.2 eV, 15.2 eV, and 18.0 eV) compare well with experimental data.
These facts can be ascribed to the four structures of the imaginary part of the
dielectric function, taking in consideration that the absorption coefficient is
directly proportional to the product of $Im ~\varepsilon (\omega)$ and the
frequency, divided by the refractive index which shows a decreasing behavior in
this energy range. The three broader structures in $\alpha$ at higher energies,
between 20 eV and 40 eV, could be also posed in correspondence with three
structures present in this energy range in the $Im ~\varepsilon (\omega)$ curve.
For these structures a slight overestimate of around 0.5\textminus1.0 eV occurs with respect to the experimental counterpart.

\section{CONCLUSIONS}
\label{sec:conclusion}
Electronic excitations and optical properties of $CdF_2$ are calculated 
within parameter-free schemes with state-of-the-art techniques. The
QP corrections within the GW approximation for the electronic self-energy of
the order of 5 eV result.
The $CdF_2$ turns out to be an indirect gap insulator with a minimum gap of
7.8 eV and a direct gap at $\Gamma$ of 8.6 eV in good accordance with
experiments. The QP spectrum of $CdF_2$ shows similarities to the $CaF_2$
spectrum (i.e., direct gap of 12.1 eV and indirect gap of 11.8 eV).
The Bethe-Salpeter scheme used for the optical absorption confirms the
existence of an exciton located about 1.0 eV below the QP gap.
The BSE spectrum for the imaginary part of the dielectric function shows an
excellent accordance with experiment.

With respect to one-particle excitations we used the GW perturbative scheme for
the electronic self-energy and calculated the energy bands and the DOS.
The electronic density of states at the gap region for $CdF_2$ and the
energy-band structure have been compared with existing data in literature.
The role of many-body effects turns out to be important in these one-particle
properties by the large opening of the energy gaps. We show, moreover, that for the optical properties, many-body
effects, treated within the BSE scheme, are fundamental to obtain a reasonable
comparison with existing experimental spectra. BSE imaginary part spectrum shows large differences in the shape which could not be recovered 
by any one-particle (GW/HSE) modified scheme at least in this material.
 
Finally, the existence of an
exciton located 1 eV below the quasiparticle gap for this compound has been discussed.
\\

\acknowledgments 

The authors acknowledge computational support provided by High Performance Computing Center
HLRS Stuttgart-Germany, and Italian SuperComputing Resource Allocation (ISCRA)-Consorzio Interuniversitario per la gestione del centro di calcolo elettronico dell`Italia Nord-orientale (CINECA) Bologna-Italy.
G.C. acknowledges the financial support of the Deutscher Akademischer Austausch Dienst
(project ref. code A/12/07710). E.C. acknowledges the financial support of Independent Development European Association IDEA-AISBL Bruxelles-Belgium. G.C. also acknowledges useful discussion with G. Malloci, A. Schroen, and E.Tosatti.


\begin{thebibliography}{99}
\bibitem{note4}
However, due to several similarities between $CdF_{2}$ and $CaF_{2}$ we expect
similar behavior also for the exciton: e.g., a Frenkel exciton as in
$LiF$\cite{fox} with a binding energy of the order of magnitude of 1.0 eV
and an exciton radius of 4\textminus5 \AA.\cite{shirley,rohlfing,albert}
\end{thebibliography}
\end{document}